\documentclass[aps,prl,groupedaddress,showpacs,amsmath,amssymb,twocolumn]{revtex4}

\usepackage{amssymb}
\usepackage{amsmath}
\usepackage{graphicx}
\usepackage{subfigure}
\usepackage{textcomp}
\usepackage{color}
\usepackage{amsfonts}
\usepackage{bbold}
\usepackage{dsfont}
\usepackage{epsfig}
\usepackage{hyperref}

\begin{document}
\title{Multiprobe quantum spin Hall bars}

\author{Awadhesh Narayan and Stefano Sanvito}
\affiliation{School of Physics, AMBER and CRANN, Trinity College, Dublin 2, Ireland}

\date{\today}

\begin{abstract}
We analyze electron transport in multiprobe quantum spin Hall (QSH) bars using the B\"{u}ttiker formalism and 
draw parallels with their quantum Hall (QH) counterparts. We find that in a QSH bar the measured resistance changes upon 
introducing side voltage probes, in contrast to the QH case. We also study four- and six-terminal geometries and derive the 
expressions for the resistances. For these our analysis is generalized from the single-channel to the multi-channel case and 
to the inclusion of backscattering originating from a constriction placed within the bar.  
\end{abstract}

\pacs{73.63.-b,73.43.-f,73.63.Hs}

\maketitle

Topological insulators are a class of materials displaying symmetry-protected topological phases, determined by time-reversal 
invariance and particle number conservation~\cite{review-kane,review-zhang,review-qsh}. A consequence of the non-trivial topology 
is the presence of edge states, which form when an interface is created with a topologically trivial material (including vacuum). In two 
dimensions a topological phase was first experimentally realized in HgTe/CdTe quantum wells. The signature of the 
edge states was first provided by two-terminal resistance~\cite{molenkamp1}, and subsequently confirmed by multi-terminal 
transport measurements~\cite{molenkamp2}. More recently, InAs/GaSb heterostructures have also been shown to have 
quantized-edge-state transport, pointing towards a quantum spin Hall (QSH) phase~\cite{rrdu}. Elemental systems such 
as Bi(111) bilayer and Sn thin films have also been predicted to be two-dimensional topological 
insulators~\cite{murakami-bi,zhang-sn}. 

A tight-binding model describing the quantum Hall effect without Landau levels was proposed first by Haldane~\cite{haldane}. 
This exhibits states moving along one direction at a given edge, the so-called \textit{chiral} edge states. Although the model does 
not require an explicit magnetic field, it breaks time-reversal symmetry. Kane and Mele then generalized this model by taking two 
copies, one for each spin, with one spin having a chiral quantum Hall effect while the other spin showing an anti-chiral quantum 
Hall effect~\cite{kanemele}. Overall the effect is that of preserving time-reversal symmetry and counter-propagating opposite 
spin \textit{helical} edge states. 

Using the concept of \textit{chiral} edge states, B\"{u}ttiker developed a picture for the quantum Hall effect in multiprobe 
devices~\cite{buttiker1}. This approach has been extensively used for analyzing experimental results, as well as gaining new 
theoretical insights. An analogous systematic study for QSH effect, which is observed in the so-called two-dimensional 
topological insulators, is still lacking. The aim of our paper is to fill this gap. 

The B\"{u}ttiker formula relating currents and voltages in a multiprobe device is~\cite{buttiker1, datta-etms}
\begin{equation}
 I_{i}=\sum_{j}(G_{ji}V_{i}-G_{ij}V_{j})=\frac{e^{2}}{h}\sum_{j}(T_{ji}V_{i}-T_{ij}V_{j})\:,
\end{equation}
where $V_{i}$ is the voltage at $i$-th terminal and $I_{i}$ is the current flowing from the same terminal. Here $T_{ij}$ is the 
transmission from the $j$-th to the $i$-th terminal and $G_{ij}$ is the associated conductance. 

For a two-terminal QSH bar the resistance can be simply written as 
$R_{12,12}=\frac{V_{1}-V_{2}}{I_{1}}=\frac{V_{1}-V_{2}}{-I_{2}}=\frac{h}{2e^{2}}$. In our notation the resistance $R_{ij,kl}$, 
is that of a setup where $i$ and $j$ are the current terminals, while $k$ and $l$ are the voltage ones. In a similar QH bar 
the measured resistance would be $R_{12,12}=\frac{h}{e^{2}}$. For simplicity here we have considered the minimum number 
of channels in both cases, namely two counter-propagating spin-polarized \textit{helical} edge states for the QSH 
bar~\cite{molenkamp3,molenkamp4}, and one \textit{chiral} edge state for the QH case. We will generalize our analysis to a 
many channels situation later, when we will also study the effects arising from backscattering in the channel. 

\begin{figure}[hb]
\begin{center}
  \includegraphics[scale=0.65]{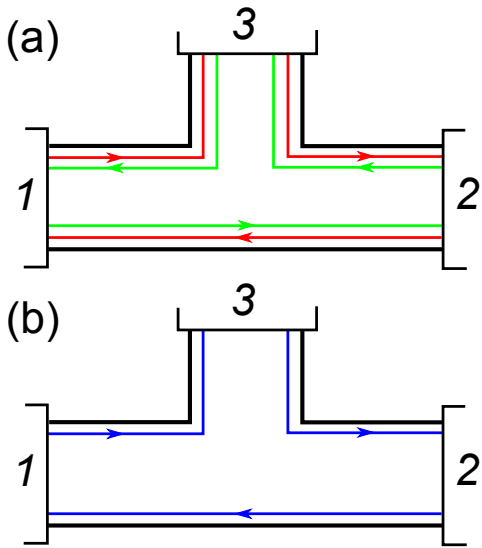}
  \caption{(Color online) Three-probe geometry for (a) a QSH and (b) a QH bar. Note that counter-propagating edge states 
(the arrows represent the direction of the electron motion) with different colors indicate the different spin directions in the QSH 
case. For a QH bar the edge states move along one direction. Here terminal $3$ is a voltage probe and draws no current.} 
\label{three-probe}
\end{center}
\end{figure}

Let us now introduce a third voltage probe (gate) in the QSH bar as shown in Fig.~\ref{three-probe}(a). The equations for 
the current derived from the  B\"{u}ttiker formula read
\begin{eqnarray}
I_{1}=\frac{e^{2}}{h}(2V_{1}-V_{2}-V_{3})\:, \nonumber \\
I_{2}=\frac{e^{2}}{h}(2V_{2}-V_{3}-V_{1})\:, \nonumber \\
I_{3}=\frac{e^{2}}{h}(2V_{3}-V_{1}-V_{2})\:.
\end{eqnarray}
We can now set $I_{3}=0$, since a voltage probe draws no current, and further assign $V_{2}=0$ as our reference potential. 
This yields $V_{3}=V_{1}/2$ and $I_{1}=\frac{3e^{2}}{2h}V_{1}$, so that the two terminal resistance becomes
\begin{equation}
\label{3QSH}
 R_{12,12}= \frac{2h}{3e^{2}}\:.
\end{equation}

Let us contrast the result just obtained with a similar three-probe analysis for a QH bar, as shown in Fig.~\ref{three-probe}(b). 
The current-voltage relations this time are
\begin{eqnarray}
I_{1}=\frac{e^{2}}{h}(V_{1}-V_{2})\:, \nonumber \\
I_{2}=\frac{e^{2}}{h}(V_{2}-V_{3})\:, \nonumber \\
I_{3}=\frac{e^{2}}{h}(V_{3}-V_{1})\:.
\end{eqnarray}
Again setting $I_{3}=0$ and $V_{2}=0$ we obtain $V_{3}=V_{1}$ and $I_{1}=\frac{e^{2}}{h}V_{1}$, hence the two-terminal 
resistance in this case is 
\begin{equation}
\label{3QH}
 R_{12,12}= \frac{h}{e^{2}}\:.
\end{equation}

When one compare equations (\ref{3QSH}) and (\ref{3QH}) with their two-terminal counterparts, an important difference 
immediately emerges. In the case of a QH bar there is no change in two-probe resistance originating from the addition of a
gate probe. This is a consequence of the voltage probe floating to the same potential as that of terminal 1, namely $V_{3}=V_{1}$.
In contrast for the QSH case, the presence of a gate increases the two-probe resistance by a factor $\frac{4}{3}$. This time 
$V_{3}$ floats to the value of $V_{1}/2$ producing the additional resistance. Such different sensitivity to a gate voltage 
is a unique observable, which distinguishes the two quantum states.

\begin{figure}[h]
\begin{center}
  \includegraphics[scale=0.60]{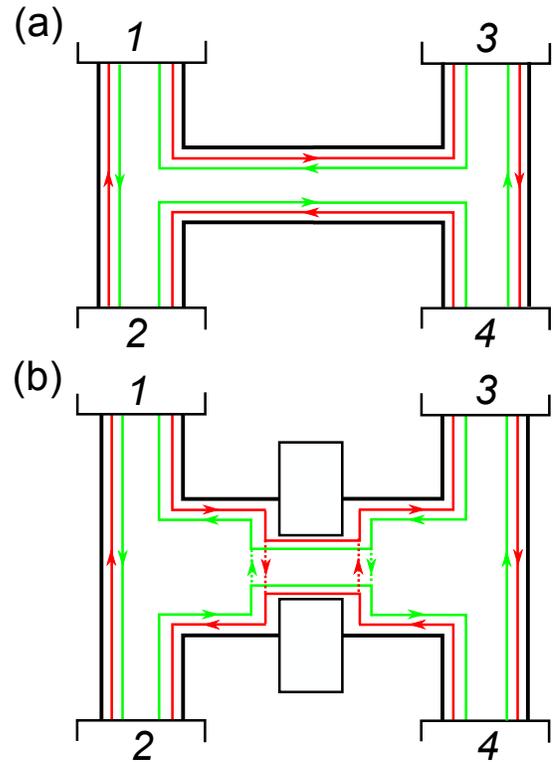}
  \caption{(Color online) (a) QSH bar in a four probe geometry and (b) the same device with now a constriction in the channel. 
  The dotted lines indicate backscattering, which, due to the topological protection, occurs only between same spin 
  channels. } \label{four-probe}
\end{center}
\end{figure}

We now turn our attention to the four-probe geometry shown in Fig.~\ref{four-probe}(a). Again it is straightforward to write 
down the current-voltage equations from the B\"{u}ttiker formula. Let us first choose terminals $3$ and $4$ as voltage probes 
($I_{3}=I_{4}=0$) and set $V_{2}=0$. This gives us $V_{3}=2V_{1}/3$, $V_{4}=V_{1}/3$ and $I_{1}=(4e^{2}/3h)V_{1}$. The 
two possible four-terminal resistances are then obtained as
\begin{equation}
 R_{12,34}= \frac{V_{3}-V_{4}}{I_{1}}=\frac{h}{4e^{2}}, \quad R_{12,12}= \frac{V_{1}-V_{2}}{I_{1}}=\frac{3h}{4e^{2}}\:.
\end{equation}
A second possibility is that of choosing the terminals $2$ and $3$ as voltage probes, so that we can set $I_{2}=I_{3}=0$. 
If we select terminal $4$ as our reference ($V_{4}=0$), we obtain $V_{2}=V_{1}/2$, $V_{3}=V_{1}/2$ and $I_{1}=(e^{2}/h)V_{1}$. 
The four-terminal resistances then read
\begin{equation}
 R_{14,23}= \frac{V_{2}-V_{3}}{I_{1}}=0,\quad R_{14,14}= \frac{V_{1}-V_{4}}{I_{1}}=\frac{h}{e^{2}}\:.
\end{equation}

Again we compare our finding with an analogous QH four-terminal bar. Using the current-voltage equations and 
substituting $I_{3}=I_{4}=0$, while choosing $V_{2}=0$, we derive $V_{3}=V_{4}=V_{1}$ and $I_{1}=(e^{2}/h)V_{1}$. 
This gives 
\begin{equation}
 R_{12,34}= \frac{V_{3}-V_{4}}{I_{1}}=0, \quad R_{12,12}= \frac{V_{1}-V_{2}}{I_{1}}=\frac{h}{e^{2}}.
\end{equation}
In contrast, choosing terminals $2$ and $3$ as voltage probes and setting $V_{4}=0$, yields $V_{2}=V_{4}=0$, $V_{3}=V_{1}$ 
and $I_{1}=(e^{2}/h)V_{1}$, with resulting resistances being
\begin{equation}
 R_{14,23}= \frac{V_{2}-V_{3}}{I_{1}}=\frac{h}{e^{2}},\quad R_{14,14}= \frac{V_{1}-V_{4}}{I_{1}}=\frac{h}{e^{2}}.
\end{equation}

Again there are important qualitative differences between the two cases. For instance, for a four-probe QH bar the local resistance,
$R_{ij,ij}$, is identical regardless of the position of the electrodes over the bar. In contrast for the QSH situation this is 
different depending on which are the electrodes which measure the current.

We now generalize the formalism to a multi-channel case and study the possible effects arising from backscattering in the 
channel. This is introduced in terms of a constriction, which narrows down the Hall bar width. Such geometry may be achieved by 
applying a split magnetic gates at the sides of the sample, as illustrated schematically in Fig. 2(b). Changing the magneitzation 
direction of the gate allows controlling the transmission across the constriction~\cite{sanvito-andreev}. Let there 
be a total of $M$ right-propagating and $M$ left-propagating channels at the edge of the QSH bar. This situation may be 
realized in Bi(111) bilayer ribbons, which host a total of six bands crossing the Fermi level~\cite{murakami-bi}. If $N$ 
channels can propagate through the constriction, then we can define the fraction of channels undergoing backscattering,
$p=\frac{M-N}{M}$. The conductance matrix in this case is written as
\begin{equation}
G_{ij}=-\frac{Me^{2}}{h}\begin{pmatrix}
        -2 & (1+p) & (1-p) & 0 \\
        (1+p) & -2 & 0 & (1-p) \\
        (1-p) & 0 & -2 & (1+p) \\ 
        0 & (1-p) & (1+p) & -2 \\
        \end{pmatrix}\:,
\end{equation}
and current-voltage relations become
\begin{eqnarray}
I_{1} &=& \frac{Me^{2}}{h}[2V_{1}-V_{2}(1+p)-V_{3}(1-p)]\:, \nonumber \\
I_{2} &=& \frac{Me^{2}}{h}[2V_{2}-V_{1}(1+p)-V_{4}(1-p)]\:, \nonumber \\
I_{3} &=& \frac{Me^{2}}{h}[2V_{3}-V_{4}(1+p)-V_{1}(1-p)]\:, \nonumber \\
I_{4} &=& \frac{Me^{2}}{h}[2V_{4}-V_{3}(1+p)-V_{2}(1-p)]\:.
\end{eqnarray}

Substituting $I_{3}=I_{4}=0$ and choosing $V_{2}=0$, we derive $V_{3}=2V_{1}/(3+p), V_{4}=(1+p)V_{1}/(3+p)$ 
and $I_{1}=4[(1+p)/(3+p)](Me^{2}/h)V_{1}$, giving
\begin{eqnarray}
R_{12,34} &=& \frac{V_{3}-V_{4}}{I_{1}}=\frac{1-p}{1+p}\frac{h}{4Me^{2}}\:, \nonumber \\
R_{12,12} &=& \frac{V_{1}-V_{2}}{I_{1}}=\frac{3+p}{1+p}\frac{h}{4Me^{2}}\:.
\end{eqnarray}

Again consider the analogous QH device, namely a four-probe bar with a constriction. The conductance matrix in this 
case can be written as
\begin{equation}
G_{ij}=-\frac{Me^{2}}{h}\begin{pmatrix}
        -1 & 1 & 0 & 0 \\
        p & -1 & 0 & (1-p) \\
        (1-p) & 0 & -1 & p \\ 
        0 & 0 & 1 & -1 \\
        \end{pmatrix}\:.
\end{equation}
Note that $G_{QH}+G_{QH}^{\dagger}=G_{QSH}$, which reminds us that the QSH state may be considered as the sum 
of a QH state and its time-reversed partner. Note also that $G_{QSH}=G_{QSH}^{\dagger}$, which is a consequence of 
time-reversal symmetry. The current-voltage relations are
\begin{eqnarray}
I_{1} &=& \frac{Me^{2}}{h}(V_{1}-V_{2}), \nonumber \\
I_{2} &=& \frac{Me^{2}}{h}[V_{2}-V_{1}p-V_{4}(1-p)], \nonumber \\
I_{3} &=& \frac{Me^{2}}{h}[V_{3}-V_{4}p-V_{1}(1-p)], \nonumber \\
I_{4} &=& \frac{Me^{2}}{h}(V_{4}-V_{3}).
\end{eqnarray}
Substituting $I_{3}=I_{4}=0$ and choosing $V_{2}=0$, we establish the relations $V_{3}=V_{4}=V_{1}=$ and 
$I_{1}=(Me^{2}/h)V_{1}$. This gives 
\begin{eqnarray}\label{R4PQH}
R_{12,34} &=& \frac{V_{3}-V_{4}}{I_{1}}=0\:,\nonumber \\
R_{12,12} &=& \frac{V_{1}-V_{2}}{I_{1}}=\frac{h}{Me^{2}}\:.
\end{eqnarray}
Equation (\ref{R4PQH}) returns us the important result that, since the voltage probes again float to the same potential 
as that of terminal 1, the resistances do not depend on $p$, namely they are not affected by the degree of backscattering. 

\begin{figure}[hb]
\begin{center}
  \includegraphics[scale=0.30,clip=true]{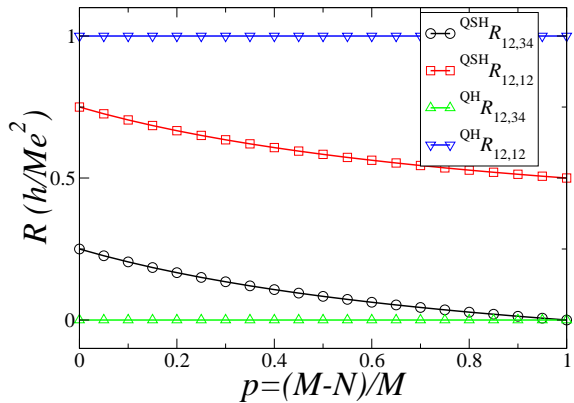}
  \caption{(Color online) Resistances of a multi-channel four-probe bar plotted as a function of the scattering parameter $p$, 
  for the QSH ($^{QSH}R$) and the QH ($^{QH}R$) cases. Increasing the backscattering results in a decrease in the resistance 
  for the QSH bar. In contrast, the QH resistances are unaffected by the backscattering.} \label{four-probe-p}
\end{center}
\end{figure}

The situation is however different for the QSH case and the differences can be appreciated by looking at Fig.~\ref{four-probe-p}, 
where the above calculated resistances are plotted as a function of the backscattering parameter $p$. For the QH case 
the resistances are independent of $p$, as explained before. In the QSH case, however, there is a decrease in resistance 
as the backscattering is increased. As $p\rightarrow 1$, we recover the result for the two-terminal geometry as terminals 3 and 4 
are decoupled from terminals 1 and 2. The resistance then returns back to the value calculated for that setup, i.e. 
$R_{12,12}=h/2Me^{2}$. In contrast, $R_{12,34}\rightarrow 0$ in the same limit, as terminals 3 and 4 float to the same 
potential.

\begin{figure}[h]
\begin{center}
  \includegraphics[scale=0.45]{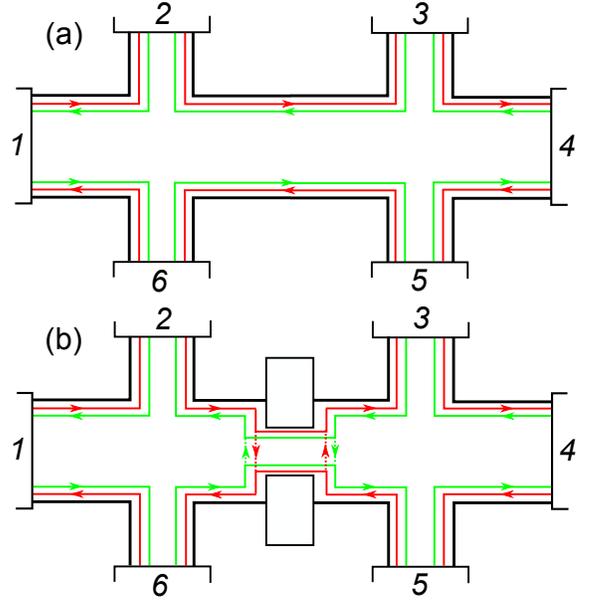}
  \caption{(Color online) (a) A QSH bar with a six-probe geometry. Here terminals $2$, $3$, $5$ and $6$ are considered 
  as voltage probes. (b) The same six probe geometry with a constriction in the channel. Because of the helical states, we consider 
  only the backscattering, which occurs between channels with the same spin.} \label{six-probe}
\end{center}
\end{figure}

Finally we continue our analysis by considering the six-terminal device sketched in Fig.~\ref{six-probe}(a). 
We choose terminals $2$, $3$, $5$ and $6$ as the voltage probes, as it is usual in such devices. Consequently we can 
set $I_{2}=I_{3}=I_{5}=I_{6}=0$ and select terminal $4$ as the voltage reference, $V_{4}=0$. Using the B\"{u}ttiker formula, 
these choices yield $V_{2}=V_{6}=2V_{1}/3$, $V_{3}=V_{5}=V_{1}/3$, and $I_{1}=(2e^{2}/3h)V_{1}$. Then resistances 
can be evaluated as
\begin{equation}
 R_{14,23}= \frac{V_{2}-V_{3}}{I_{1}}=\frac{h}{2e^{2}},\quad R_{14,14}= \frac{V_{1}-V_{4}}{I_{1}}=\frac{3h}{2e^{2}}\:.
\end{equation}

As before, we continue to draw parallels with the corresponding QH bar. Carrying on by substituting $I_{2}=I_{3}=I_{5}=I_{6}=0$, 
and $V_{4}=0$ into the current-voltage relations, we calculate $V_{2}=V_{3}=V_{1}$, $V_{5}=V_{6}=V_{4}=0$ and 
$I_{1}=(e^{2}/h)V_{1}$. The voltage probes at the top edge (2 and 3) float to the potential of terminal 1, while those at 
the bottom edge (5 and 6) float to the potential of terminal 4. From these the resistances are obtained to be
\begin{equation}
 R_{14,23}= \frac{V_{2}-V_{3}}{I_{1}}=0,\quad R_{14,14}= \frac{V_{1}-V_{4}}{I_{1}}=\frac{h}{e^{2}}\:.
\end{equation}

We finally consider the effect of the constriction introduced in the channel, as shown schematically in Fig.~\ref{six-probe}(b). 
The conductance matrix can be written down as
\begin{equation}
G_{ij}=-\frac{Me^{2}}{h}\begin{pmatrix}
        -2 & 1 & 0 & 0 & 0 & 1 \\
        1 & -2 & (1-p) & 0 & 0 & p \\
        0 & (1-p) & -2 & 1 & p & 0 \\ 
        0 & 0 & 1 & -2 & 1 & 0 \\
        0 & 0 & p & 1 & -2 & (1-p) \\
        1 & p & 0 & 0 & (1-p) & -2 \\ 
       \end{pmatrix}\:.
\end{equation}
and the current-voltage relations read
\begin{eqnarray}
I_{1} &=& \frac{Me^{2}}{h}(2V_{1}-V_{2}-V_{6})\:, \nonumber \\
I_{2} &=& \frac{Me^{2}}{h}[2V_{2}-V_{1}-(1-p)V_{3}-pV_{6}]\:, \nonumber \\
I_{3} &=& \frac{Me^{2}}{h}[2V_{3}-V_{4}-(1-p)V_{2}-pV_{5}]\:, \nonumber \\
I_{4} &=& \frac{Me^{2}}{h}(2V_{4}-V_{3}-V_{5})\:, \nonumber \\
I_{5} &=& \frac{Me^{2}}{h}[2V_{5}-(1-p)V_{6}-V_{4}-pV_{3}]\:, \nonumber \\
I_{6} &=& \frac{Me^{2}}{h}[2V_{6}-(1-p)V_{5}-V_{1}-pV_{2}]\:.
\end{eqnarray}
Following now the choice of voltage probes and voltage reference as before, we derive $V_{2}=V_{6}=[(2-p)/(3-2p)]V_{1}$, 
$V_{3}=V_{5}=[(1-p)/(3-2p)]V_{1}$ and $I_{1}=[(2-2p)/(3-2p)](Me^{2}/h)V_{1}$. The resistance values are then
\begin{eqnarray}
R_{14,23} &=& \frac{V_{2}-V_{3}}{I_{1}}=\frac{1}{1-p}\frac{h}{2Me^{2}}, \nonumber \\
R_{14,14} &=& \frac{V_{1}-V_{4}}{I_{1}}=\frac{3-2p}{2-2p}\frac{h}{Me^{2}}.
\end{eqnarray}
Note that we recover the previous results for no constriction when we set $p=0$, as we should. Ultimately we repeat 
our comparison with QH bar with a constriction. 

It is again illustrative to look at the conductance matrix, 
\begin{equation}
G_{ij}=-\frac{Me^{2}}{h}\begin{pmatrix}
        -1 & 0 & 0 & 0 & 0 & 1 \\
        1 & -1 & 0 & 0 & 0 & 0 \\
        0 & (1-p) & -1 & 0 & p & 0 \\ 
        0 & 0 & 1 & -1 & 0 & 0 \\
        0 & 0 & 0 & 1 & -1 & 0 \\
        0 & p & 0 & 0 & (1-p) & -1 \\ 
       \end{pmatrix}\:,
\end{equation}
which gives us the following current-voltage relations
\begin{eqnarray}
I_{1} &=& \frac{Me^{2}}{h}(V_{1}-V_{6}), \nonumber \\
I_{2} &=& \frac{Me^{2}}{h}(V_{2}-V_{1}), \nonumber \\
I_{3} &=& \frac{Me^{2}}{h}[V_{3}-(1-p)V_{2}-pV_{5}], \nonumber \\
I_{4} &=& \frac{Me^{2}}{h}(V_{4}-V_{3}), \nonumber \\
I_{5} &=& \frac{Me^{2}}{h}(V_{5}-V_{4}), \nonumber \\
I_{6} &=& \frac{Me^{2}}{h}[V_{6}-(1-p)V_{5}-pV_{2}].
\end{eqnarray}
Employing the same voltage probes and reference conditions as in the QSH case, we obtain $V_{2}=V_{1}$, $V_{3}=(1-p)V_{1}$, 
$V_{5}=V_{4}=0$, $V_{6}=pV_{1}$ and $I_{1}=(1-p)(Me^{2}/h)V_{1}$, and the resistances
\begin{eqnarray}
R_{14,23} &=& \frac{V_{2}-V_{3}}{I_{1}}=\frac{p}{1-p}\frac{h}{Me^{2}}\:,\nonumber \\
R_{14,14} &=& \frac{V_{1}-V_{4}}{I_{1}}=\frac{1}{1-p}\frac{h}{Me^{2}}\:.
\end{eqnarray}
From the final expression for the resistances it is clear that in the case of a six-probe geometry backscattering is relevant for both
the QH and QSH case. In particular now in a QH bar, backscattering mixes the upper and lower edges and the voltages at the 
upper and lower voltage probes do not equal that of the left and right terminals, respectively. This contrasts the case of the 
four-probe geometry with constriction, in which all the voltage probes were at the same edge and as a consequence the 
resistance remained unaffected by the presence of backscattering in the channel.

\begin{figure}[htb]
\begin{center}
  \includegraphics[scale=0.30,clip=true]{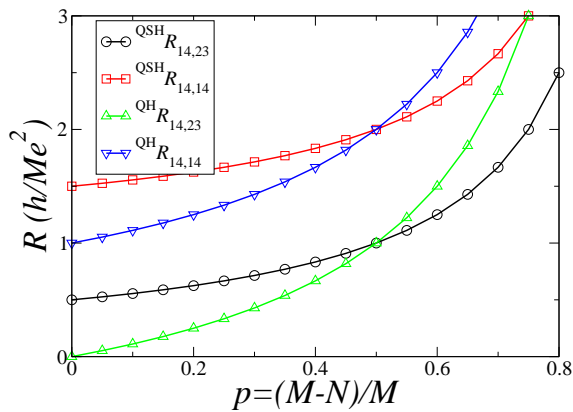}
  \caption{(Color online) Resistances plotted as a function of the backscattering $p$, for QSH ($^{QSH}R$) and QH 
  ($^{QH}R$) six terminal bars. Increasing backscattering results in an increase in the resistance for both cases.} 
  \label{six-probe-p}
\end{center}
\end{figure}

A better comparison between the QH and QSH case can be obtained by plotting the calculated resistances as a function 
of the backscattering parameter $p$, as shown in Fig.~\ref{six-probe-p}. In this case all the resistances increase with 
increasing backscattering. This is because the constriction is now placed between the terminals, where the current is
measured and this reduces the number of channels passing from terminal 1 to 4. For small $p$, the QH resistances are 
lower than their QSH counterparts. An interesting cross-over occurs at $p=0.5$, where an equal number of channels 
are backscattered and transmitted. The QH and QSH resistances become equal and thereafter the QH resistance rises 
above the QSH one.

In conclusion, we have studied multi-terminal QSH devices based on the B\"{u}ttiker approach of ballistic edge transport. 
We have derived expressions for resistances in three-, four- and six-terminal geometries and compared these to the ones 
obtained for the corresponding QH bars. Furthermore, we have analyzed the effect of backscattering, which may be 
introduced by an electrostatically-controlled constriction in the channel. For a four-probe setup the resistance change as 
a function of the backscattering has markedly different behaviour for QH and QSH bars. In the case of a six-terminal 
geometry we have found an interesting crossover occurring across the point, where half of the channels are completely
backscattered.

This work is financially supported by Irish Research Council (AN) and the European Research Council, QUEST project 
(SS). AN would like to thank Rupesh Narayan and Brajesh Narayan for discussions which inspired this work.\\

\textit{Note added}: Immediately prior to the manuscript submission a related work Ref.~\cite{chulkov-multiprobe}, appeared 
which has partial overlap with our results. In our work we have made comparisons of QSH devices with analogous QH bars and have 
made explicit the similarities and differences. The main focus of this paper is to model backscattering from constrictions 
and provide suggestions for experimental measurements.

\end{document}